\documentclass[reprint,
 amsmath,amssymb,
 aps,
]{revtex4-2}

\usepackage{graphicx}
\usepackage{dcolumn}
\usepackage{bm}
\usepackage{physics}
\usepackage[usenames,dvipsnames]{color}
\usepackage{relsize} 
\usepackage{enumitem}
\usepackage{mathptmx}
\usepackage{graphicx} 
\usepackage{float}
\usepackage[english]{babel}
\usepackage{svg}
\usepackage{subcaption}

\newcommand\moldstruct{M{\relsize{-2}OL}DS{\relsize{-2}TRUCT}}

\begin{document}

\preprint{APS/123-QED}

\title{Protein structure classification based on X-ray laser induced Coulomb explosion}
\thanks{ESI avialable at: }%

\author{Tomas André}
\author{Ibrahim Dawod}
\altaffiliation[Also at ]{European XFEL, Holzkoppel 4, DE-22869 Schenefeld, Germany.}
\author{Sebastian Cardoch}
\author{Nicu\c{s}or T\^\i mneanu}
\author{Carl Caleman}
\email{carl.caleman@physics.uu.se}
\altaffiliation[Also at ]{Center for Free-Electron Laser Science, Deutsches Elektronen-Synchrotron, Notkestraße 85 DE-22607 Hamburg, Germany.}
\affiliation{Department of Physics and Astronomy, Uppsala University, Box 516, SE-751 20 Uppsala, Sweden.}%

\author{Emiliano De Santis}
\altaffiliation[Also at ]{University of Rome Tor Vergata \& INFN, Rome, 00133, Italy.}
\affiliation{
Department of Chemistry – BMC, Uppsala University, Box 576, SE-751 23 Uppsala,
Sweden.
}%

\date{\today}

\begin{abstract}
We simulated the Coulomb explosion dynamics due to the fast ionization induced by high-intensity X-rays in six proteins that share similar atomic content and shape. We followed and projected the trajectory of the fragments onto a virtual detector, providing a unique explosion footprint. After collecting 500 explosion footprints for each protein, we utilized principal component analysis and t-distributed stochastic neighbor embedding to classify these. The results show that the classification algorithms were able to separate proteins on the basis of explosion footprints from structurally similar proteins into distinct groups. The explosion footprints, therefore, provide a unique identifier for each of the proteins. We envision that method could be used concurrently with single particle coherent imaging experiments to provide additional information on shape, mass, or conformation.
\end{abstract}

\maketitle


\section{\label{sec:level1}Introduction}

Radiation damage studies have been of interest since the idea of Single Particle Imaging (SPI) was first introduced~\cite{neutze2000potential}. SPI aims to obtain structural information from non-crystalline samples with high-intensity femtosecond duration X-ray pulses from an X-ray Free-Electron Laser (XFEL) that elastically scatter onto a detector~\cite{chapman_x-ray_2019}. So far, atomic resolution reconstruction of nanometer-sized systems such as single proteins has been hindered by several technical challenges~\cite{aquila_linac_2015}. One aspect of fundamental importance is the X-ray-induced damage that destroys the sample. Photons with energies commonly used for imaging experiments primarily photoionize, leaving atoms in excited electronic states that, within femtoseconds, decay radiatively and non-radiatively. Free electrons originating from these events cause additional changes to the electronic configuration through collision. In small proteins, this secondary damage is not as severe since free electrons with a mean free path greater than the particle's size escape, leaving behind charged ions ~\cite{caleman2011feasibility}. The excess positive charge buildup leads to significant electrostatic forces that break the structure apart in a process known as Coulomb explosion.

\begin{figure}[!h]
    \centering
    \includegraphics[width=0.75\linewidth]{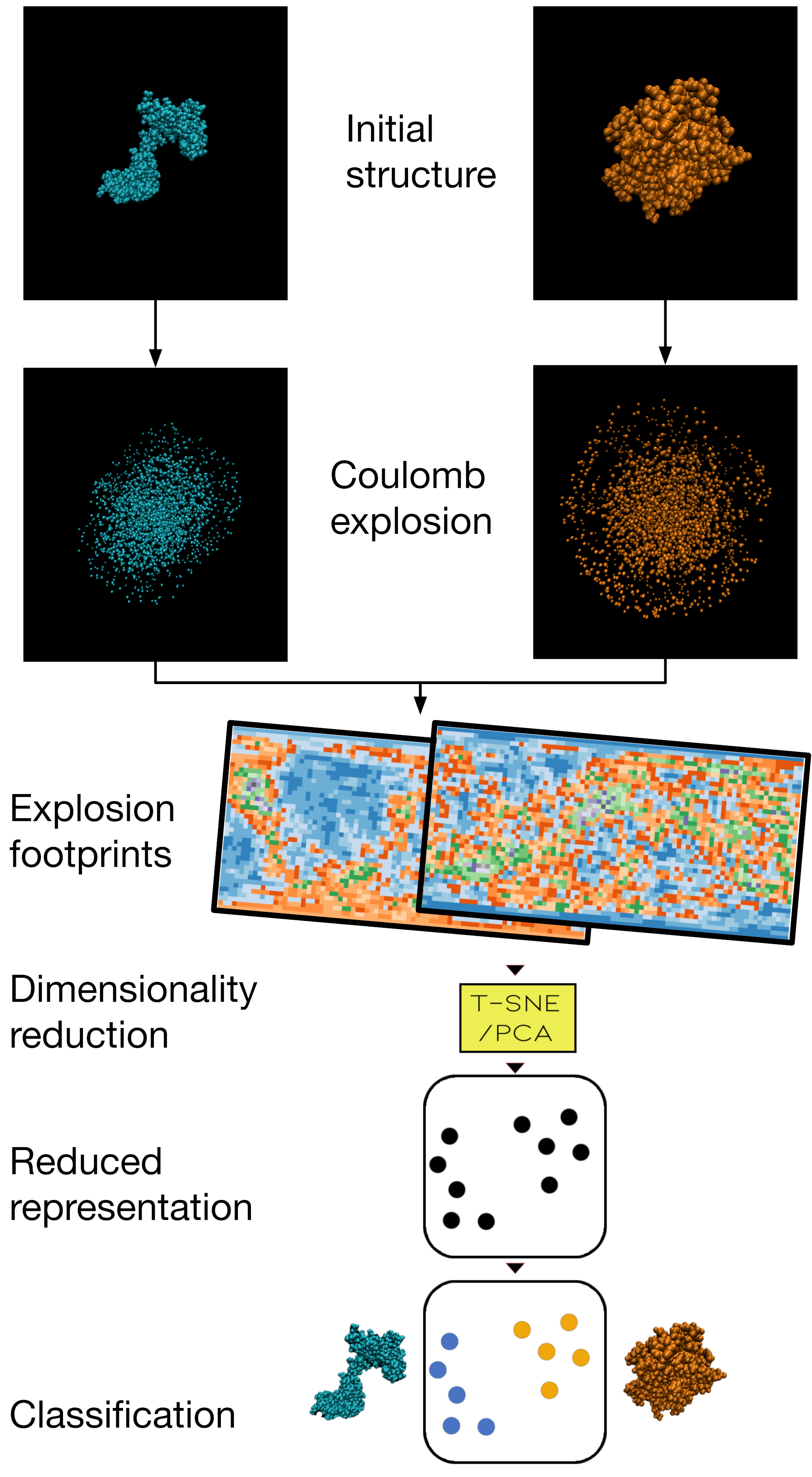}
    \caption{Conceptual overview of the process. First, the system is simulated under the exposure of an XFEL pulse using the MC/MD code, \moldstruct. Afterwards, we create the explosion footprints from the trajectories. The data are then fed through a dimensionality reduction funnel, making the high-dimensional footprints more comparable. We can then show the reduced footprints as points in a 2D space. Lastly, in this 2D space we can employ clustering to classify the data.}
    \label{fig:diagram}
\end{figure}

Coulomb explosion imaging is a single-molecule structural determination technique of a sample stripped of its electrons, that traces the fragments by measuring the momenta of the resulting ions in coincidence~\cite{vager_coulomb_1989}. XFELs provide a tool to carry out Coulomb explosion imaging since X-rays can be tuned to target specific inner shells while reaching highly-charged states via sequential single-photon absorption~\cite{kukk_molecular_2017,takanashi_ultrafast_2017,boll_x-ray_2022}. \"Ostlin {\it et al.}~\cite{ostlin2018} simulated X-ray-induced Coulomb explosions on lysozyme to construct time-integrated \textit{explosion footprints} generated by projecting carbon and sulphur ions trajectories onto a virtual detector and concluded these maps could be used to determine the protein's orientation during exposure. Our present study takes this idea further by classifying explosion footprints on three pairs of proteins that share stoichiometric and conformational similarities. Unlike conventional Coulomb explosion images, the explosion footprints used in this study are constructed uniquely from ion trajectories and carry no coincidence or momentum information.

In this simulation study, we aim to answer the following question: {\it In an XFEL experiment, is it feasible to separate structurally similar proteins solely based on the explosion footprint?} To do this 
we model the interaction between the X-ray laser and single proteins using a Monte Carlo/Molecular Dynamics (MC/MD) code, similar to~\cite{dawod_moldstruct_2024}, that computes electronic occupation and ion dynamics. After tracing the ions' trajectory, we carry out a dimensionality reduction to project explosion footprints in two dimensions to assess if sufficient structural information is preserved to uniquely separate explosion footprints from similar proteins. A schematic summarizing our work is presented in Fig.~\ref{fig:diagram}. We see this study as a first step to develop a technique that can capture additional complementary information during SPI experiments to aid orientation recovery algorithms such as expansion-maximization-compression needed for reconstruction~\cite{august2024,loh_reconstruction_2009}.

\section{Results}
We begin by preparing the simulation environment, for the systems outlined in Table~\ref{tab:systems}. To quantify the similarity of the three selected pairs of proteins, we use the local Distance Difference Test (lDDT)~\cite{lddt}, calculated using~\cite{waterhouse2024}. lDDT is a superposition-free score which evaluates local distance differences in a model compared to a reference structure. The lDDT values span from 0 to 1, where 1 corresponds to perfect structural match. 

\begin{figure*}
    \centering
    \includegraphics[width=\linewidth]{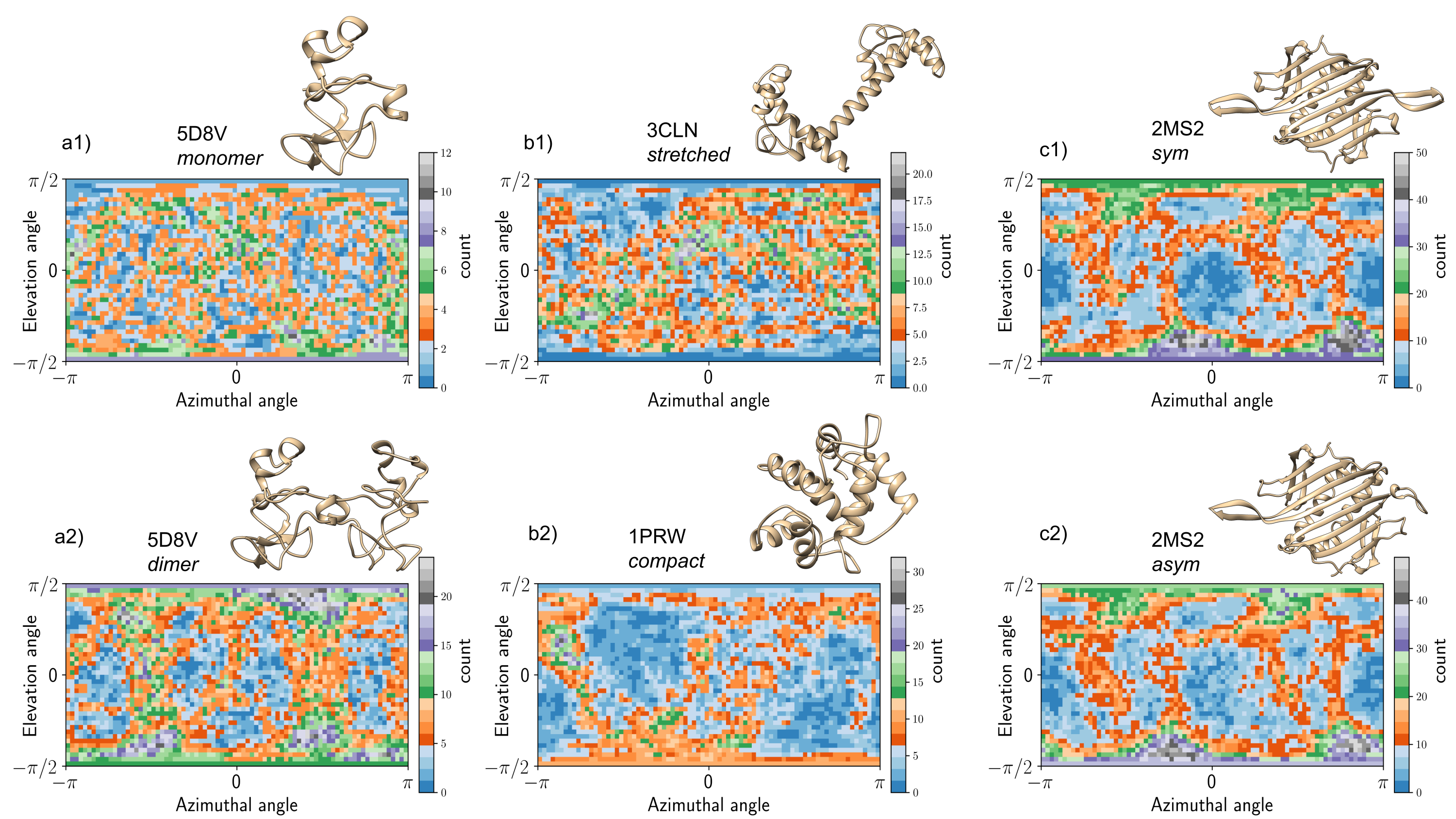}
    \caption{Visual representation of systems together with an explosion footprint from a singular explosion, labeled after the different cases studied. Three pairs of systems were selected for this study. First  monomer a1)  and a dimer a2) of a High-Potential Iron-sulfur Protein. The second pair was two systems with similar atom count but different structure: b1) a stretched and b2) a compact folding of a calmodulin protein. The last pair is the one we expect to be hardest to separate: 
    c1) a symmetric and c2) asymmetric folding of the MS2 virus coat protein.}
    \label{fig:systems}
\end{figure*}

Each protein is placed in a vacuum at a fixed orientation, and  after a standard equilibration procedure, we acquire snapshots of the structure at distinct time steps to perform MC/MD simulations. Model details are available in the SI. The Coulomb explosions are triggered by a temporal Gaussian-shaped X-ray pulse with a \(10\)~fs full width at half maximum duration, \(600\)~eV photon energy, and $5\times10^{6}$~photons/nm$^2$ fluence. We perform \(500\) simulations for every protein that follow the electron occupation and ion dynamics as a function of time. After 100~fs from the exposure, we project the resulting ion trajectories onto a unit sphere using the direction of the unit velocity vector. See the SI for additional details. To visualize the explosion footprints, the spherical signal is distorted into two dimensions (equirectangular projection), akin to some world maps, resembling a theoretical full spatial area detector, with the x- and y-axes representing azimuthal and elevation angles, respectively. Examples of these two-dimensional footprints are shown in Fig. \ref{fig:systems}. The averages of many footprints originating from the same protein can easily be distinguished by eye. However, singular footprints exhibit high variance due to differences in the protein structure at the moment of exposure and the inherent probabilistic nature of photon-matter interaction.

\begin{table*}
    \centering
    \begin{tabular}{|c|c|c|c|c|c|c|} \hline 
        Systems &\multicolumn{2}{c|}{HiPIP~\cite{hirano2016}}&  \multicolumn{2}{c|}{Calmodulin~\cite{babu1988structure,calmodulin03}}& \multicolumn{2}{c|}{MS2 coat protein~\cite{2MS2}}                  \\       \hline 
        PDB-name & 5D8V& 5D8V & 1PRW & 3CLN & 2MS2 & 2MS2                 \\       \hline 
        Alias & Dimer & Monomer &  Compact & Stretched & Sym & Asym           \\       \hline 
        Atoms    &2430   & 1215 & 2184 & 2240 & 3858 & 3858    
        \\ \hline
        lDDT & \multicolumn{2}{c|}{0.38} & \multicolumn{2}{c|}{0.77} & \multicolumn{2}{c|}{0.89} 
        \\       \hline 
    \end{tabular}
    \caption{Information about the three different systems we investigate, the HiPIP dimer/monomer~\cite{hirano2016}, the stretched/compact calmodulin~\cite{babu1988structure,calmodulin03} proteins and the symmetric and asymmetric dimers of the MS2 virus~\cite{2MS2}. We list the names used by the PDB-database, the alias we will refer to them as and the number of atoms (hydrogen atoms included) in each protein. We also list the pairwise lDDT score~\cite{lddt} between the proteins in the system.  For a visual representation see Fig. \ref{fig:systems}.} 
    \label{tab:systems}
\end{table*}

\begin{figure}
    \centering
    \includegraphics[width=0.9\linewidth]{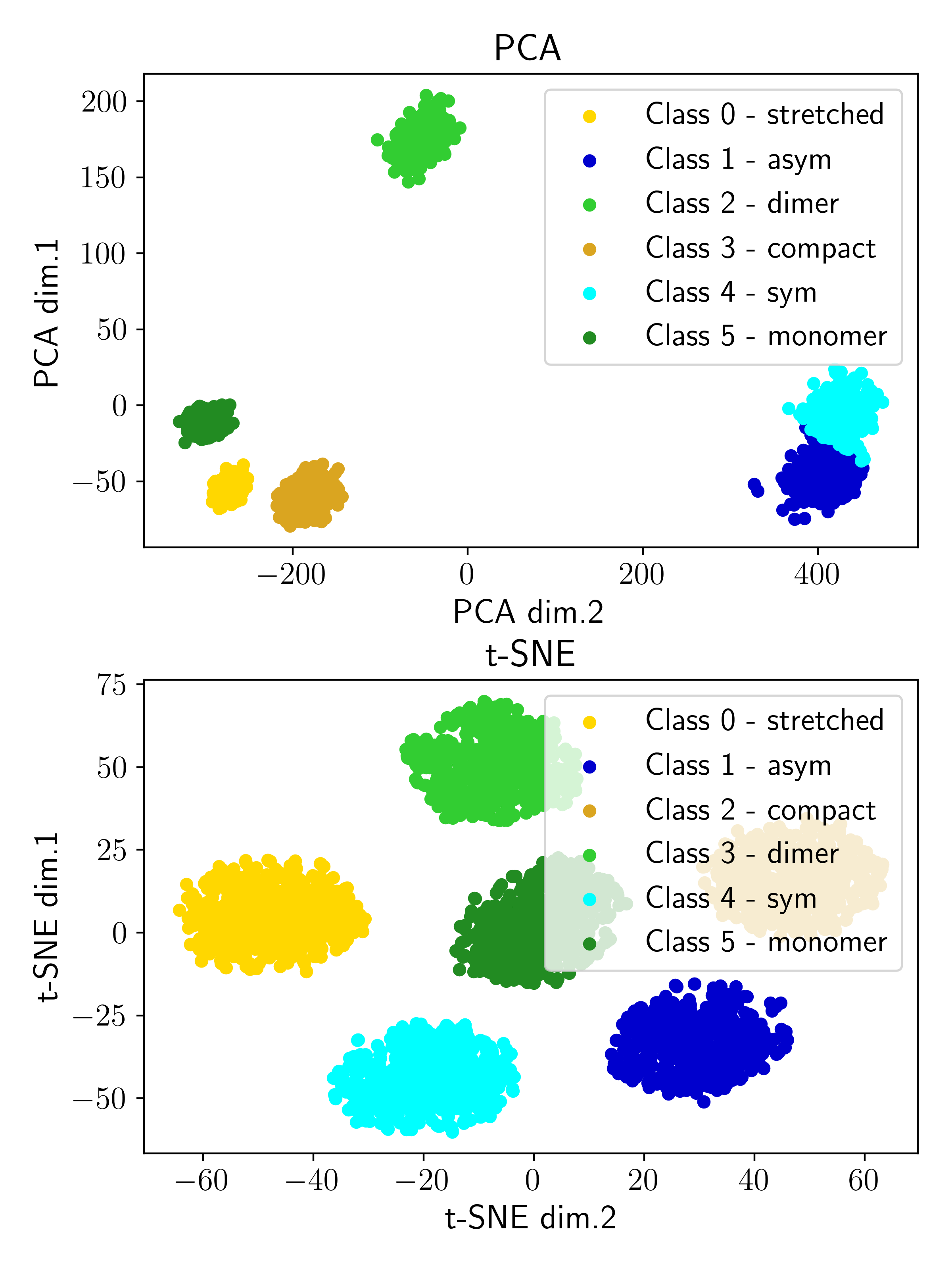}
    \caption{Scatter plots of the protein explosion footprints in the reduced space, with dimensionality reduction by PCA~\cite{PCA} (top) and t-SNE~\cite{van2008tsne} (bottom). The clustering is evaluated with the adjusted random score, where PCA gives a score of 0.97 and t-SNE achieves the maximum score of 1.00, a perfect match. The original orientation of the proteins was kept fixed and all proteins can be  classified in distinct classes. The classes are predicted by k-means and labeled to what system they mainly correspond to.} 
    \label{fig:clusters}
\end{figure}

To classify the footprints on an individual basis, we make use of principal component analysis (PCA) ~\cite{PCA} and t-distributed stochastic neighbor embedding (t-SNE)~\cite{van2008tsne} to reduce the dimensionality of each explosion footprint to two dimensions. We use both techniques since they highlight different aspects of the problem. The t-SNE method produces a reduced space much more suitable for clustering algorithms, i.e. well separated clusters of similar sizes. The PCA method preserves distances, making it easier to compare systems. The clustering itself is done using k-means, a simple clustering algorithm that groups similar data points together by finding the best centers for each group and iteratively adjusting these centers until the groups are as compact and distinct as possible.

The PCA and t-SNE scatter plots of the explosion footprints in the reduced spaces are shown in Fig.~\ref{fig:clusters}.
Both algorithms can easily separate the monomer and dimer of the same protein, depicted as dark and light green dots. The algorithms are also able to separate two of the same protein in two different conformations, with one being compact and the other stretched, shown as dark and light yellow dots. Comparing how PCA places the clusters, we note that PCA seems to regard the monomer more similar to the compact and the stretched structures, which is interesting since this is not obvious to the human eye looking at the footprints, Fig.~\ref{fig:systems}. The final and most astonishing result is that t-SNE is able to separate asymmetric and symmetric structures, depicted as light and dark blue dots. These two proteins have the same amino acid sequence and their structures are almost identical. The main differences between them are their distinct FG loops. For the symmetric structure, all FG loops are well defined $\beta$-hairpins, while for the asymmetric structure one of the FG loops is collapsed towards the main protein body \cite{BRODMERKEL2022}. Despite this very subtle difference, t-SNE clearly separates the two structures, while PCA is not capable of doing so.

To evaluate the quality in the clustering we employ the adjusted random score (ARS)~\cite{ARS}, a measure to compare the similarity of 
two sets of clusters, on a scale between 1 and -1. A value of 1 indicates perfect agreements between the clusters, 0 indicates similarities between clusters are random, and -1 indicates perfect disagreement. By setting one set of clusters to the correct values and the other set of clusters to the predicted values, we can use the ARS as a metric for how well the predicted clusters fit the correct clusters. By computing the ARS of the k-means predicted clusters and the true-labels, we find the PCA achieves a value of 0.97 and t-SNE a value of 1.00. This means that t-SNE are able to group individual explosion footprints together with perfect precision, at least in our study. 

It should be kept in mind that these footprints are not normalized, therefore information about number of atom and by extension their approximate mass (due to a positive correlation between number atoms and atomic mass in the proteins we study) is encoded via the intensity variations. In an attempt to estimate whether the algorithms would work even if we removed the information about the number of atoms in the proteins, we normalized all the integrated intensities in the footprints to one. We note that we achieve similar ARS values and reduced spaces using our normalized footprints (results not shown) as in Fig.~\ref{fig:clusters}, thus the clustering is not exclusively measuring the number of atoms, which would essentially just be counting the mass. 

The results presented so far assume we know the orientation of the protein, which is typically not the case for SPI experiments done today. However,  attempts to pre-orient the proteins with external electric fields exist~\cite{kadek2021flash}, and earlier studies indicate that the orientation can be retrieved from the explosions~\cite{ostlin2018}. In addition, in an SPI experiment where the diffracion image is recorded simultaneously, these can be used to find the orientation~\cite{loh_reconstruction_2009}. We attempt to distinguish explosion footprints using t-SNE without any knowledge about the orientation and find the only two systems that are possible to separate are the dimer and the monomer (results not shown).

\begin{figure}
    \centering
    \includegraphics[width=\linewidth]{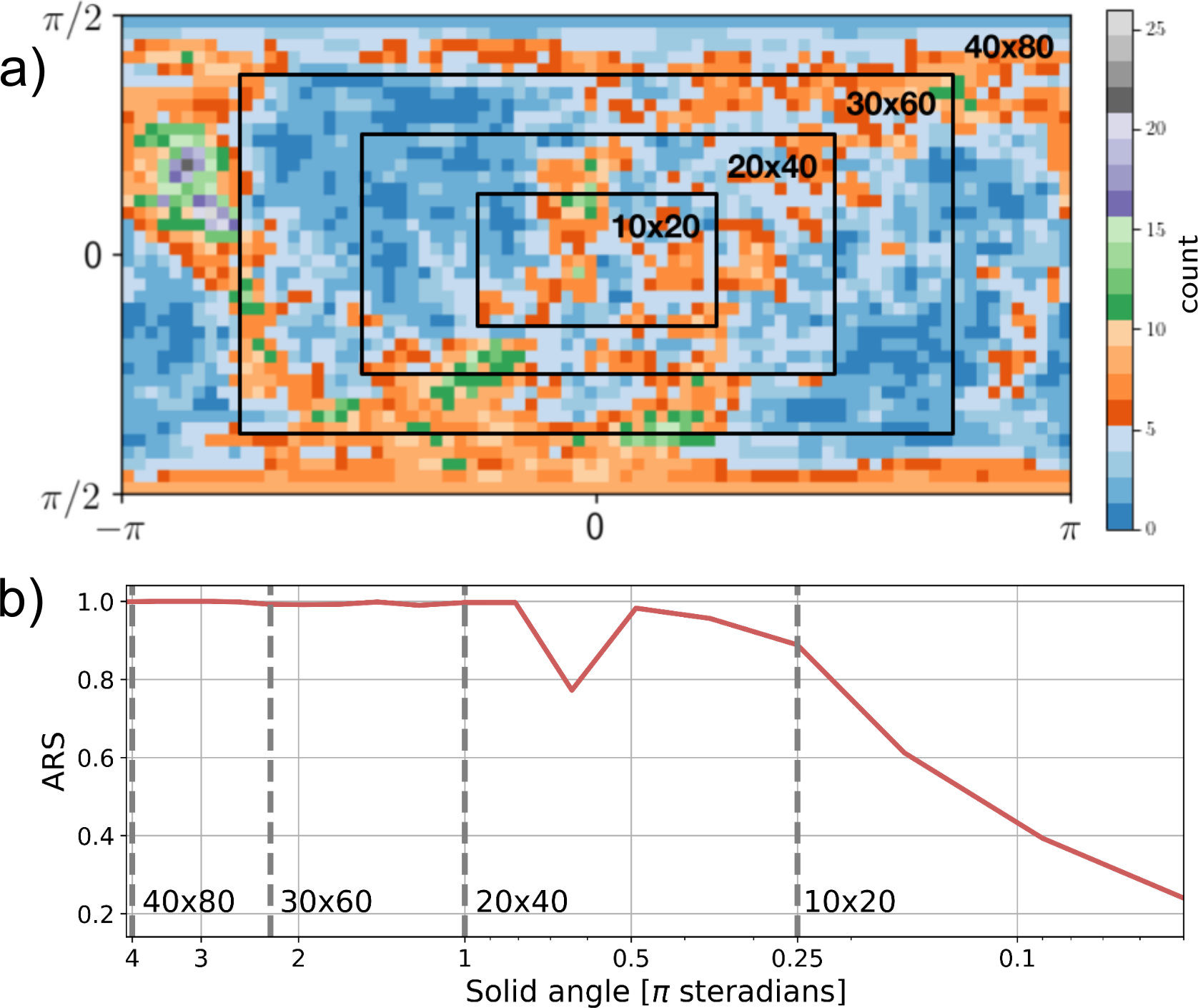}
    \caption{\textbf{a)} Explosion footprint showing different detector sizes. \textbf{b)} Adjusted random score based on the reduced t-SNE spaces compared between different sizes of detectors. Each pixel covers an average solid angle of 0.004~steradians. For reference, a score of \(0.8\text{--}0.9\) corresponds to a good clustering in this case.}
    \label{fig:robustness}
\end{figure}

So far in our study we considered the usage of a spherical \(4\pi\)-detector, which is an idealization that is not feasible using current experimental setups. By removing  the outer pixels of the explosion footprints we can reduce the solid average coverage and approach something more similar to a planar detector as seen in Fig. \ref{fig:robustness}a. To gauge how this impacts the clustering we calculate the ARS from clustering the reduced t-SNE spaces while incrementally trimming the edges of the image, effectively utilizing only a central portion of the detector while maintaining the original solid angle per pixel resolution. We present the dependence in Fig. \ref{fig:robustness}b. We see that the clustering remains effective even for smaller planar-like detectors, with a detector area that would be feasible to cover experimentally.

\section{Discussion and Conclusion}

We have used a Monte Carlo/Molecular Dynamics model to simulate Coulomb explosions of 6 different proteins initiated by ultrafast X-ray pulses from X-ray Free Electron Lasers. By mimicking a full-spatial detector we can extract the directions of the ions, from which we can create a unique explosion footprint. By using the explosion data from proteins with known orientation, we have been able to accurately classify all of the different proteins across all degrees of structural variations. We also demonstrated the robustness of the classification for detectors covering smaller solid angles~\ref{fig:robustness}b.
Based on the assumption that proteins exposed to an intense X-ray pulse explode in a reproducible manner, we have investigated the possibility to use the information contained in the explosion to separate proteins based on structure. This is the first step towards finding ways to extract more detailed structural information from measuring the direction of ions ejected from proteins. 

t-SNE was able to classify and differentiate all the proteins we tried, from the less complex cases of the monomer/dimer to the higher complex case of sym/asym. Our study shows that the algorithm seems to be able to resolve shape, and atom count to a great extent as well as working with smaller detectors. The most striking is that it could separate the sym and asym structures, which have identical amino acid chains, and very similar structures.  These two structure exhibit an lDDT score of 0.89, as seen in Table 1, which is comparable to the level of accuracy that the machine learning folding predictor Alphafold can predict protein backbone structure~\cite{jumper2021highly}. 
Thus our presented method can distingush proteins roughly on the same level as Alphafold can predict structures. This is promising, but we can also conclude that none of the algorithms we used, PCA and t-SNE, were able to separate the proteins if we did not provide the information about the orientation. Protein orientation might be possible to determine either by physically orienting the proteins using external fields~\cite{Marklund2017a} or by measuring the trajectories of tag ions in the protein structures, like sulfur~\cite{ostlin2018}.  Earlier simulation studies have suggested that adding a thin layer of water around the proteins would both improve the heterogenity between individual proteins~\cite{Patriksson2007a,water_juncheng,mandl}, and make the explosion footprints more well defined~\cite{ostlin2018} than in this study.

To put the numbers of our results into context we can hypothesize an experiment to see how it compares to our simulations. Utilizing a position-sensitive microchannel plate detector~\cite{KORKULU2023232} with a 120 mm diameter with a sample-detector distance of 120 mm, which is an achievable sample-detector distance at the SPB/SFX endstation at EuXFEL~\cite{mancuso2019single}, we calculate that such detector would cover a solid angle of $0.25\pi$~steradians. Compared to the resolution of our simulations, this case corresponds to using only the $10\times20$ central pixels (see Fig.~\ref{fig:robustness}a). Fig.~\ref{fig:robustness}b gives an indication of the ARS for any size of detector. We highlight the solid angle value of $0.25\pi$ in the  Fig.~\ref{fig:robustness}b to easily compare a realistic detector geometry to our simulations. 

The chances to retrieve structural information from the explosions would most likely improve if one employed more advanced machine learning algorithms than used here. However such algorithms often require training data, which could be complicated to generate experimentally. To achieve high resolution structures from SPI measurements, it would be beneficial to combine the explosion footprints with the diffraction images, and maybe even with an X-ray spectrometer to monitor the atomic processes caused by the ionization in the sample. Even if the explosion footprints in themselves could not give high resolution structures, they could provide information about global parameters, such as mass and shape of the protein, which possibly can be used as support for phasing algorithms.

This is a simulation study, and what we describe will not be trivial to investigate experimentally. However, based on our findings we believe that efforts towards structural classification of proteins based on Coulomb explosions is an interesting path to improve single particle imaging using XFEL. If coupled with machine learning folding predictors like AlphaFold~\cite{jumper2021highly}, it could even be a step towards determine the structure based on machine learning and explosion footprints only, without the need of the X-ray diffraction. Protein explosion could in principle be achieved by tabletop femtosecond lasers, which are much more accessible than XFELs.  

\bibliography{ref}

\end{document}


\preprint{APS/123-QED}

\title{Supplemental Information -- \\Protein structure classification based on X-ray laser induced Coulomb explosion}

\author{Tomas André}
\author{Ibrahim Dawod}
\altaffiliation[Also at ]{European XFEL, Holzkoppel 4, DE-22869 Schenefeld, Germany.}
\author{Sebastian Cardoch}
\author{Nicu\c{s}or T\^\i mneanu}
\author{Carl Caleman}
\email{carl.caleman@physics.uu.se}
\altaffiliation[Also at ]{Center for Free-Electron Laser Science, Deutsches Elektronen-Synchrotron, Notkestraße 85 DE-22607 Hamburg, Germany.}
\affiliation{Department of Physics and Astronomy, Uppsala University, Box 516, SE-751 20 Uppsala, Sweden.}%

\author{Emiliano De Santis}
\altaffiliation[Also at ]{University of Rome Tor Vergata \& INFN, Rome, 00133, Italy.}
\affiliation{
Department of Chemistry – BMC, Uppsala University, Box 576, SE-751 23 Uppsala,
Sweden.
}%

\date{\today}
\maketitle


\section{Hybrid Monte Carlo/Molecular Dynamics model}
Based on our previous version of \moldstruct, designed for larger systems such as protein crystals and liquids~\cite{dawod_moldstruct_2024}, we have made a modified version specifically to simulate single particles. The previous version has been benchmarked against experimental data~\cite{nass2020structural,beyerlein2018ultrafast}. 
Our hybrid MC/MD is available at Github\footnote{\moldstruct~builds on GROMACS and is available at \url{https://github.com/moldstruct/mc-md/}} along with a user guide. 
The model builds on the same spirit as other hybrid approaches~\cite{kozlov2020hybrid,ho2017large,jurek2016xmdyn}. The interaction between photons and atoms can induce changes in the electronic occupation through different transitions. We include photo-ionization, fluorescence and Auger-Meitner decay. The model assumes all photons and electrons escape and, therefore, does not include electron collision effects. The model also omits shake-up and shake-off processes. The positions of the atoms are propagated using classical MD implemented in the software GROMACS~\cite{GROMACS}, where bonded and non-bonded interactions are dynamically altered based on the level of ionization of the atoms. Within a single MD time step \(\Delta t\), the program calculates changes to the electronic occupations by evaluating the timescales \(\Delta\tau_{l}\) of each transition \(l\) using the following iterative MC algorithm.
\begin{enumerate}
    \item Determine all rates \(R\) corresponding to transitions between electronic states given the current X-ray pulse conditions.
    \item Looping over all atoms in the system.
    \begin{enumerate}
    \item Compute the timescale $\Delta \tau$ for all transition using a random number $r_{\tau}\in [0, 1]$ sampled for each transition from a Poisson distribution with mean $\lambda = R \Delta \tau$ using the equation
    \begin{equation}
        \Delta \tau = -\frac{\log(r_{\tau})}{R}\ .
    \end{equation}
    \item While \(\sum_{l}\Delta\tau_l < \Delta t\)
    \begin{enumerate}
\item The transition \(l\) with the smallest \(\Delta\tau\) modifies the electronic populations, if it is smaller than \(\Delta t\).
\end{enumerate}
\end{enumerate}
\item Return the charge of all atoms to modify interactions and carry out MD.
\end{enumerate}
The rates for flouresence and Auger-Meitner decay can all be calculated beforehand while the photo-ionization rate requires scaling by the current intensity of the pulse. 
The atomic data (energy levels and transitions rates) is taken from the collisional-radiative code CRETIN~\cite{scott2001cretin}, but any code that can compute similar quantities could be used - such as XATOM~\cite{jurek2016xmdyn}, FLYCHK~\cite{chung2005flychk} or FAC~\cite{gu2008}. We have used CRETIN with a screened hydrogenic model for levels distinguished by principal quantum number $N$. In our calculations we only include the first three principal quantum numbers $K, L$ and $M$.  

After the Monte Carlo computation, we determine the possibility of charge transfer between two atoms. We have previously shown using ab-initio simulations that highly charged amino-acids stabilize when the electron from the hydrogen atom transfers to a nearby charged heavier atom, the hydrogen attains a charge and disassociates from the molecule~\cite{Aminosyror}, carrying a positive charge. By excluding charge transfer, the hydrogen atom in our model will rarely acquire a net-charge, due to its low photo-ionization cross section. To account for this, we use the classical over-barrier model (COB). Previous experiments and simulations have shown that this model can be used to capture this process~\cite{schnorr2014electron, boll_x-ray_2022, ho2023x}. The model is based on a two-body interaction, where the donor atom with net-charge $Q_\text{D}$ at position $\textbf{R}_\text{D}$ can transfer an electron with binding energy $E_\text{p}$ to an acceptor atom with charge $Q_\text{D} \leq Q_\text{A}$ at position $\textbf{R}_\text{A}$. The electron will feel the Coulomb potential from the donor atom with effective charge $Q_\text{D} + 1$ and the acceptor atom $Q_\text{A}$. This potential will at some critical distance $R_\text{c}$ provide the electron with a potential larger than the barrier between the two atoms, which allows for charge transfer. It is calculated as
\begin{equation}
    R_\text{c} = \frac{Q_\text{D} + 1 + \sqrt{(Q_\text{D}+1)Q_\text{A}}}{|E_\text{p}|}.
\end{equation}
When the distance between the two atoms is $R\leq R_\text{c}$, we change the electronic occupations of the two atoms accordingly. For each time step, charge transfer is computed after the photon-matter calculation by the MC algorithm has ended. We iterate through the atoms and determine whether the charge transfer occurs. Since charge transfer changes the electronic states of the system, the order at which the atoms are iterated through could influence the calculations. To avoid any systematic bias we shuffle the indices of the atoms for each MD time step.

\section{Protein explosion simulations}
For the pairs of proteins, in order of increasing complexity, we consider
\begin{itemize}
    \item A dimer/monomer system of a high potential iron-suflur protein (HiPIP)~\cite{hirano2016}. The systems have distinct shape and atom count, while having similar distribution of atomic content.
    \item Two different foldings of a Calmodulin protein, one stretched out~\cite{babu1988structure} and one compact~\cite{calmodulin03}. The systems have distinct shape, while having similar atom count, and atomic content.
    \item Two dimers of the MS2 virus coat-protein, a symmetric C/C dimer and an asymmetric A/B dimer~\cite{2MS2}. The systems have similar shape, atom count, and atomic content.

\end{itemize}
These 3 pairs are chosen since they display different type of similarities, and ideally we can distinguish all types of structural differences. The six systems are illustrated in Fig. \ref{fig:systems}. 
All MD simulations are carried out in GROMACS 4.5.6~\cite{GROMACS}, and the explosion simulations with the hybrid modifications described earlier, \moldstruct. After energy minimization using steepest decent, we equilibrate the system at constant temperature (300~K) and pressure (1~bar), using 1~fs time step, while keeping the protein's orientation fixed. From the end this simulation, we sample structural variations in the protein that serve as an initial starting point for the MC/MD simulations. This is done for a total of 500 simulations per system. For the MC/MD simulations we assume particle injection speed is negligible compared to the speed of the ions during the explosion, as typical injection speeds are on orders of $10^1$-$10^2$~m/s while the ions in our simulations are ejected at $10^{4}$-$10^{5}$~m/s  The MC/MD simulations are carried out with a time step of 1~as. To assure that our systems have fully exploded, we monitor the ratio of potential energy to total energy of the system. When this ratio is less than 1\% we assume that all relevant dynamics have occurred and that the atoms are merely expanding into space. All simulations reached this threshold in under 100~fs. All of the MD simulations are initialized using the CHARMM36 forcefield~\cite{klauda2010update} with a modification to handle the iron-sulfur clusters of the HiPIP~\cite{da2022development}. The simulated XFEL pulse is Gaussian-shaped with a peak at $t=20$~fs. The pulse has a fluence of $5\cdot10^{6}$~photons/nm$^2$ and photon energy 600~eV with a full width half maximum of 10~fs. The pulse parameters are chosen to resamble those at the small quantum systems (SQS) beam line at the European XFEL. 

\section{Analysis of explosion footprints}
The analysis builds on tracking the direction of atoms after an XFEL induced explosion, as the location where the different atoms end up should encode information about the geometry of the exploded system. To imitate such a 3D detector that can track the direction of the atoms we adopt the same binning methodology used in an early study~\cite{ostlin2018}. A spherical envelope of radius $R=1$ is created around the protein, with uniformly distributed bin-centers along the azimuthal and elevation angles. The bins themselves take the form of spherical caps and their size is determined by the orthodomic radius, denoted $r$. For the $j$'th ion the distance to the $i$'th bin-center is given by 
\begin{equation}
    d_{ij} = \cos^{-1}\left(\hat{\bm{x}}_i \cdot \hat{\bm{v}}_j   \right) R,
\end{equation}
where $\hat{\bm{x}}_i$ is the position of the $i$'th bin-center, and $\hat{\bm{v}}_j$ the unit velocity of the $j$'th ion, both in Cartesian coordinates. Since both vectors are normalized, the arc length has the same magnitude as the angle between the two vectors. Ion $j$ then belongs to all bins where $d_{ij} < r$ is fulfilled, adding +1 each bin for each ion belonging to it. Consistent with our early study~\cite{ostlin2018} we chose $r=0.1$. We opt to use 40 elevation bins and 80 aziumuthal bins. The resulting bins can then be viewed as 2D heatmaps, where each bin is a pixel with a value corresponding to the number of ions ejected in that direction, illustrated as explosion footprints shown in Fig. \ref{fig:systems}. With the given number of bins each pixel would cover an average solid angle of $\Omega = 0.004$~steradians. The number of bins and $r$ can be seen as  parameters since more highly resolved data could always be redcuded into larger bins by pooling pixels together.

\section{Dimensionality reduction}
We can consider each explosion footprint as a single point in an \(N\)-dimensional space, where each dimension corresponds to a pixel in the image. Given our binning choice, we have a $40\times80$-dimensional space. Comparing such high-dimensional data directly can be challenging, and we may miss underlying patterns or groupings.
To overcome this issue, we employ dimensionality reduction with principle component analysis (PCA)~\cite{PCA} as well as a technique known as t-distributed stochastic neighbor embedding (t-SNE) ~\cite{van2008tsne}. These methods, which we use through the \textit{sklearn} python module~\cite{scikit-learn}, allow us to transform the high-dimensional data into a lower-dimensional space while preserving as much information as possible, we choose to reduce the data down to 2D, drastically simplifying the problem. 
Both methods accomplish the same result in different ways and therefore highlight different features. PCA reduces the data by projecting it onto the two largest components of the covariance matrix, with the idea that most of the variation within the data is captured in there. t-SNE map points that are close together in a high dimensional tend to be close in the lower dimensional space. The data is projected into a 2D space, then, using gradient decsent, points are moved around in the lower dimensional space to minimize the Kullback-Leiber divergence~\cite{kullback}, a measure on how good a probability distribution is for approximating another one. By minimizing this measure we guarantee that the lower dimensional space loses minimal information compared to the higher dimensional space.
This results in two distinct types of reduce spaces. The PCA space conserves distances and can therefore more easily gauge how similar or distinct two footprints are. t-SNE on the other hand produces a much more homogeneous space, however is better at grouping similar data together.

Ideally, if explosions from the same system exhibit similar characteristics, they will be clustered together in the lower dimensional space. Conversely, explosions from differing systems should be positioned farther apart, making it easier to distinguish and identify distinct clusters corresponding to each system. Plotting the reduced dimensions in a 2D scatter plot provides an effective way to assess similarities and spread between explosion footprints.

\bibliography{ref}